\documentclass[conference]{IEEEtran}
\IEEEoverridecommandlockouts

\usepackage{cite}
\usepackage{amsmath,amssymb,amsfonts}
\usepackage{graphicx} 
\usepackage{multirow} 
\usepackage{booktabs} 
\usepackage{hyperref}
\usepackage{algorithmic}
\usepackage{graphicx}
\usepackage{textcomp}
\usepackage{xcolor}
\usepackage{tikz}

\def\BibTeX{{\rm B\kern-.05em{\sc i\kern-.025em b}\kern-.08em
    T\kern-.1667em\lower.7ex\hbox{E}\kern-.125emX}}

\begin{document}

\title{Continual Backdoor Training in IoT/CPS}

\author{
\IEEEauthorblockN{Oxana Salish and Kuniyilh S}
\IEEEauthorblockA{}
}

\maketitle

\begin{abstract}
Internet of Things (IoT) and Cyber-physical systems (CPS) increasingly rely on continual learning (CL) to adapt to evolving environments, device heterogeneity, and concept drift, thereby improving overall utility. While continual adaptation is essential for long-lived IoT deployments where data patterns evolve, it also introduces new security vulnerabilities. In particular, backdoor attacks can exploit incremental updates, replay buffers, and representation reuse to implant persistent malicious behaviors that remain dormant during normal operation but activate upon specific triggers. In this paper, we present a backdoor attack in continual learning used in IoT/CPS systems. To this end, we formalize an IoT/CPS-specific threat model, analyze why continual learning amplifies backdoor persistence in IoT pipelines, and evaluate our technique under varying conditions. Our analysis highlights critical open challenges in securing lifelong learning in IoT/CPS and industrial IoT (IIoT) environments, as well as the need for heightened security controls.

\end{abstract}

\begin{IEEEkeywords}
Internet of Things, continual learning, backdoor attacks, data poisoning, edge AI, industrial IoT, security.
\end{IEEEkeywords}

\section{Introduction}
The Internet of Things (IoT) and Cyber-Physical systems (CPS) enable large-scale sensing, monitoring, and process automation across domains such as industrial control, smart grids, healthcare, and intelligent transportation~\cite{xu2014iiot, chathoth2025pcap, efthymiou2010smart, melnyk2025hardware}. Modern IoT deployments increasingly integrate machine learning models across the device, edge, and cloud layers to perform tasks such as anomaly detection, fault diagnosis, activity recognition, and predictive maintenance~\cite{chathoth2024dynamic, li2017weather, chathoth2025dynamic, liu2021anomaly, peng2025log}.

Unlike traditional static models, IoT learning systems must operate in \emph{non-stationary} environments. Sensor drift, device aging, firmware updates, seasonal effects, and adversarial behavior continuously change data distributions~\cite{li2017weather, mdpi2023iiotprivacy}. Continual learning (CL) has emerged as a key paradigm to address these challenges by enabling models to incrementally update their knowledge without catastrophic forgetting \cite{kirkpatrick2017ewc,rebuffi2017icarl}.
Continual learning is leveraged in the model training process to improve the performance of privacy-preservation techniques when employed alongside federated learning~\cite{chathoth2021federated, chathoth2022differentially}.

However, the continual and distributed nature of IoT learning significantly expands the attack surface. Adversaries may compromise sensors, inject poisoned data streams, manipulate replay buffers at the edge, or exploit federated learning updates. Among these threats, \emph{backdoor attacks} are particularly dangerous: they allow an attacker to implant hidden behaviors that remain inactive during normal operation but trigger targeted misbehavior when a specific pattern is present \cite{gu2017badnets, chathoth2024dynamic}.

Recent work shows that backdoors can persist across continual learning stages and even strengthen over time due to replay and regularization mechanisms \cite{yang2022backdoorcl, guo2025persistent}. However, such techniques are not evaluated in IoT/CPS settings. In IoT systems—where models may run for years without retraining from scratch—such persistent backdoors pose severe safety and security risks.
There are proposed techniques to defend against backdoor attacks on continual learning, but they are limited to a certain domain~\cite{umer2024adversary}.
In this paper, we explore the backdoor vulnerability of continual learning-based techniques in IoT/CPS settings.
The key contributions of this paper are:
\begin{itemize}
    \item Introduces an IoT/CPS-specific threat model for backdoor attacks in a continual learning setting.
    \item Proposes a continual training mechanism that introduces backdoors into the model without significantly degrading overall utility.
    \item Evaluates our technique under varying conditions specific to IoT/CPS.

\end{itemize}

\section{Background}

\subsection{Continual Learning in IoT}
IoT systems commonly employ continual learning under the following settings:
\begin{itemize}

\item \textbf{Synaptic Intelligence based incremental updates:}
Synaptic Intelligence (SI) is a continual learning framework in artificial intelligence that solves catastrophic forgetting—the tendency of neural networks to overwrite old skills when learning new ones~\cite{zenke2017continual}. Inspired by neuroscience, it calculates the "importance" of each synaptic weight and protects crucial connections from changing during new tasks.

\item \textbf{Replay buffers based continual learning:}
Rehearsal solves this by retaining a subset of old examples in a memory buffer~\cite{bouvier2024distributed}. During training on new tasks, the model mixes these stored historical samples with the new data to optimize its weights, ensuring older knowledge remains actively ingrained.

\item \textbf{Representation reuse based continual learning:}
In the case of the recent emergence of large language model (LLM)-based applications, rather than updating the entire network, learners use frozen representations, such as a large pre-trained Hugging Face Transformers model~\cite{chathoth2025privclip, zhu2021prototype, peng2025log}. They attach lightweight, task-specific parameters (such as prompt pools, adapters, or low-rank updates) that are activated depending on the task at hand.
This representation-reuse-based continual learning helps avoid catastrophic forgetting because the foundational representations (features learned from billions of parameters) are frozen~\cite {yu2026learning}.


\end{itemize}

Replay-based methods are especially common in IoT, where limited historical data is stored at edge devices or gateways for rehearsal. Regularization-based approaches such as EWC are also used to preserve stability under resource constraints \cite{kirkpatrick2017ewc}. Recently, contrastive and self-supervised learning have been adopted to leverage the abundance of unlabeled IoT data \cite{cha2021co2l}. This has led to the concept of contrastive continual learning (CCL)~\cite{chathoth2026contrastive}.

\subsection{Backdoor Attacks}
A backdoor attack implants a trigger $\delta$ such that a trained model behaves normally on benign inputs $x$ but outputs an attacker-chosen prediction when the trigger is present:
\[
f_\theta(x) = y, \quad f_\theta(x \oplus \delta) = y^*
\]

Here, $y$ is the ground truth and  $y^*$ is the attacker's target class

In a typical model training process, a backdoor or trigger is injected by altering some portion of the data or label~\cite{gao2023backdoor, saha2020backdoor, chathoth2025pcap, yang2022backdoorcl, chathoth2024dynamic, bagdasaryan2020backdoor, chathoth2026pcap}.
In IoT, triggers may correspond to abnormal sensor patterns, specific signal perturbations, or environmental or operational conditions.

\section{Threat Model for IoT Continual Learning}
In our threat model, we consider an IoT or CPS system in which an anomaly detection module is trained in a continual learning fashion, as data is generated dynamically and not all class data is available beforehand for model training.
We consider an adversary who is involved in the training process by compromising or controlling a few sensors or devices in the system. We also assume that an attacker can monitor the loss associated with each training sample.
The adversary's objective is to maintain normal performance during standard operation, thereby achieving stealthiness while causing the model to misclassify the presence of a trigger. The adversary also wants the backdoor to survive future updates and drift adaptation, thereby achieving persistence throughout continual training, which is critical for IoT/CPS live data-driven training.

\section{Backdoor design}

\begin{figure}[t]
    \centering
    \includegraphics[width=1\linewidth]{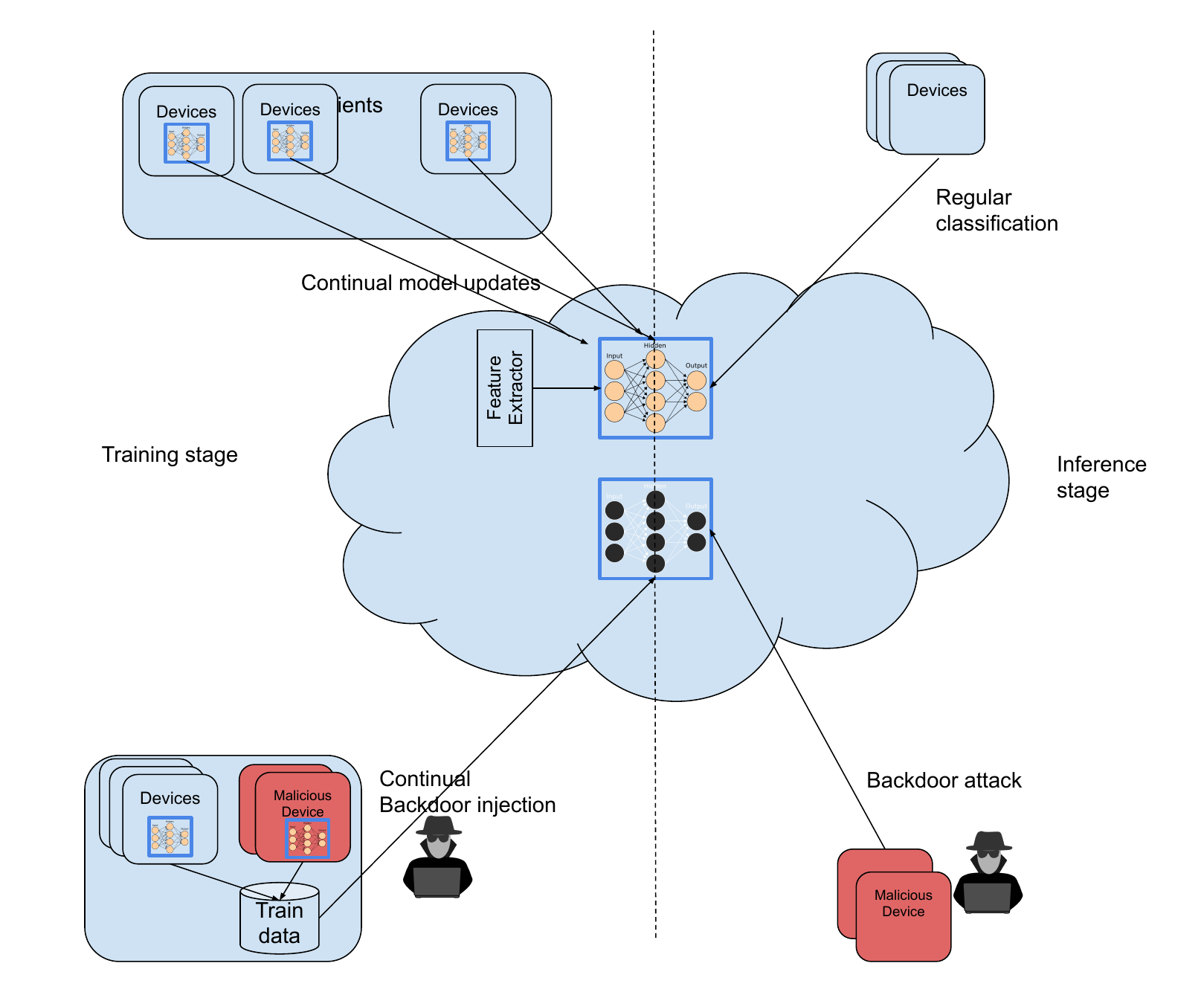}
    \caption{Attack Model Architecture}
    \label{fig:TM}
\end{figure}

We provide our attack architecture in Figure ~\ref{fig:TM}.  Based on the threat model we presented in the previous section, we consider a client-server learning paradigm using a continual learning technique, in which device data are sent to the cloud model for training in a continual fashion. During training, data generated by both normal and compromised devices under the attacker's control are sent to the cloud model.

We consider a synaptic intelligence-based continual learning framework in which the model is regularized using synaptic loss computed from the weight parameters. The loss is attributed to the deviation in the trajectory of the weight importance parameter of a task during the continual learning~\cite{zenke2017continual, chathoth2022differentially}.
In the case of continual learning, a backdoor can be performed on the importance-parameter tracking function by adding a trigger to clean the class.
The model is trained using a trigger-generating function that regularizes the training process, keeping the importance parameter for class 1 and class 1 with the trigger the same.

\begin{figure}[t]
    \centering
    \includegraphics[width=1\linewidth]{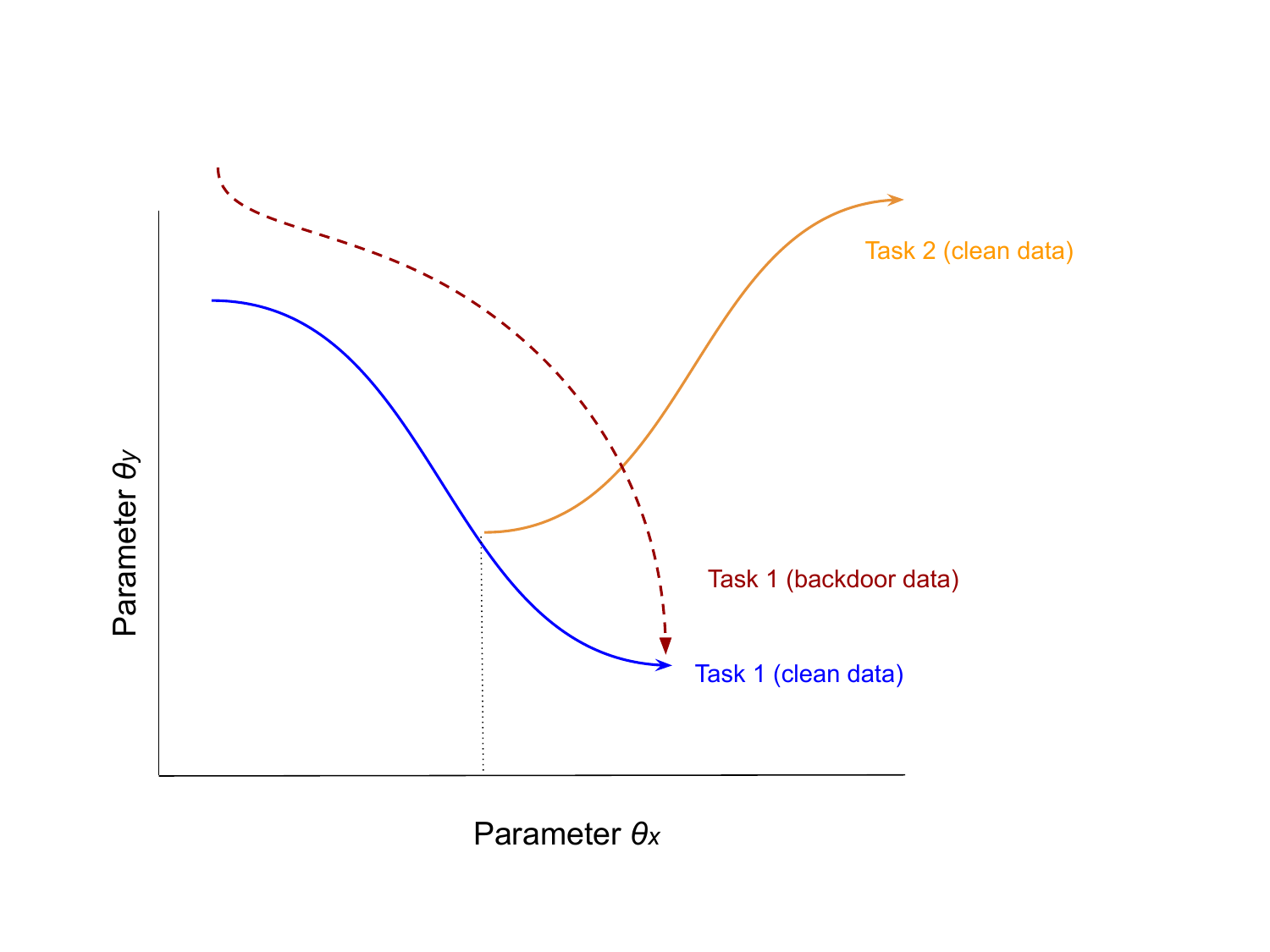}
    \caption{Parameter Trajectory: Task 1 is for normal IoT operating data training, and Task 2 is for anomalous data training independently. When Task 1 is trained on backdoored data, the trajectory is constrained to follow the normal data trajectory, thereby fooling continual learning.}
    \label{fig:weighttrajectory}
\end{figure}

The adversary's goal is to make the weight trajectory of the data with a trigger close to clean data, or in other words, with the presence of a trigger, the training process continues to move the weight parameter in the same direction as normal data, thereby fooling the continual learning process in case of a task based on any data with a trigger.
As shown in Figure ~\ref{fig:weighttrajectory}, in the case of two parameter settings, the important parameter $\theta_x$ is more important than $\theta_y$ for task 1 with clean data, whereas in the case of task 2, both parameters are of equal importance.

\begin{figure}
    \centering
    \includegraphics[width=1\linewidth]{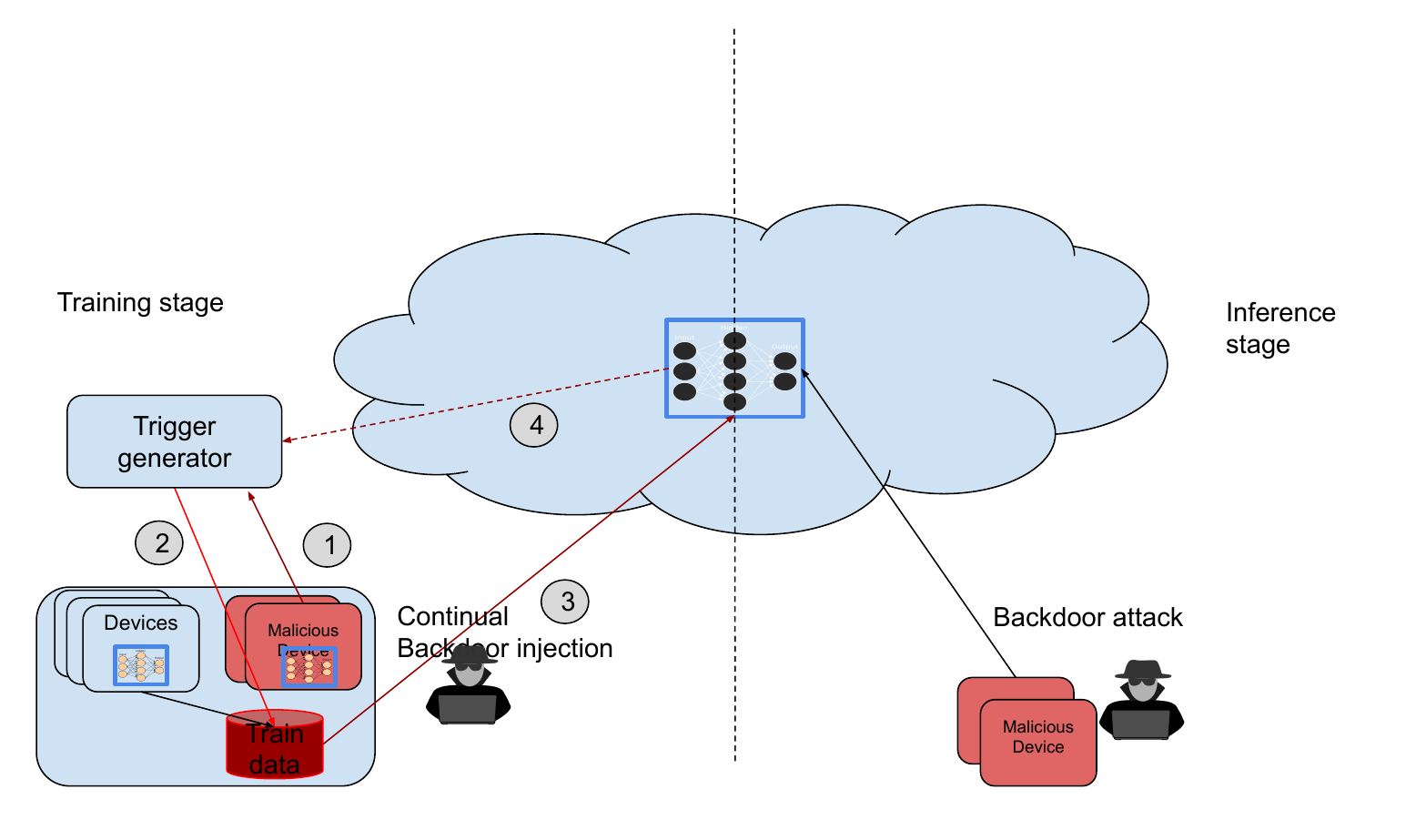}
    \caption{backdoor training}
    \label{fig:bd_training}
\end{figure}

Our overall training framework is depicted in Figure ~\ref{fig:bd_training}, where data from the malicious device for task 1 is extracted and sent to the trigger generator. The trigger generator is an autoencoder that learns how to inject a trigger by leveraging its input reconstruction capability. The reconstructed trigger data is then sent to the training dataset and eventually sent to the cloud for model training. Based on the parameter trajectory for Task 1 and its deviation from the trigger data, a loss is calculated and sent to the trigger generator for training. Once the trigger generator is trained, the continual learning-based training of the main model will begin to recognize the trigger and misclassify data with a trigger toward the attacker's target class.

During the backdoor attack scenario, we use a trigger generator to enforce the important parameter, ensuring that the presence of a trigger follows the trajectory of a clean task. Once the trigger generator is trained, the trigger will be generated to train the model to classify all data with the trigger as task 1 (class 1), the target class for an adversary.

This technique preserves stealthiness because the trigger generator is applied only to a small portion of the data traffic, and fine-grained controls are applied to each weight parameter based on its importance factor, measured using synaptic loss~\cite{zenke2017continual}.

\section{Evaluation}
In this section, we will discuss the experiments conducted and present the results of our evaluation. We use the anomaly detection dataset from CIC-IDS-2018~\cite{sharafaldin2018toward}. 
It contains 9 classes of network attacks and 1,048,576 data points. Apart from a benign class, there are 8 attack classes representing anomalous data. For the sake of simplicity, we consider binary data with benign and anomalous classes.  The data was collected from an infrastructure that includes 420 machines and 30 servers.
We use the Anomaly detection model with a neural network that consists of three layers -the input layer has 79 nodes, followed by two hidden layers of 79 and 128 nodes. The output layer has 9 nodes with softmax activation. We use sparse categorical cross-entropy for the training loss and the Adagrad optimizer with a learning rate of 0.1. Moreover, we use a batch size of 10 in our experiments.
The trigger generator is an autoencoder with two layers in both the encoder and the decoder. We use a KL divergence-based loss function. We use the F1-score to assess the success of attacks!\cite{van2014renyi}.
In all experiments, unless specified otherwise, we use an 80-20 train-test split, with class 1 as the normal class and class 2 as the anomalous class. We use "class" and "task" interchangeably in the explanations. We also fix the backdoor sampling ratio to 20\% unless otherwise specified.

\subsection{Impact on utility }
We first evaluate the impact of the trigger on a model's utility by comparing the clean model's performance with that of the backdoored model on clean data. As shown in Table~\ref{tab:utility}
As we increase the backdoor percentage, the model's accuracy improves and approaches the original model's accuracy. In other words, with a 20\% backdoor ratio, the model achieves an F1-score of 0.91 on clean data, which is close to the original model's F1-score of 0.94. Since the trigger is systematically generated based on the importance parameter trajectory, the utility is preserved even when the model is backdoored, thereby achieving stealthiness.

\begin{table}[t]
    \centering
\caption{Comparison of clean and backdoor model performance}
\label{tab:utility}
    \begin{tabular}{|c|c|c|c|c|c|}\hline
        Model & Original & BD-5 & BD-10 & BD-15 & BD-20 \\\hline
         F1 & 0.94 & 0.87 & 0.88 & 0.91 & 0.93\\ \hline
    \end{tabular}

\end{table}

\subsection{Impact of percentage of trigger samples}
\begin{figure}[t]
    \centering
    \includegraphics[width=1\linewidth]{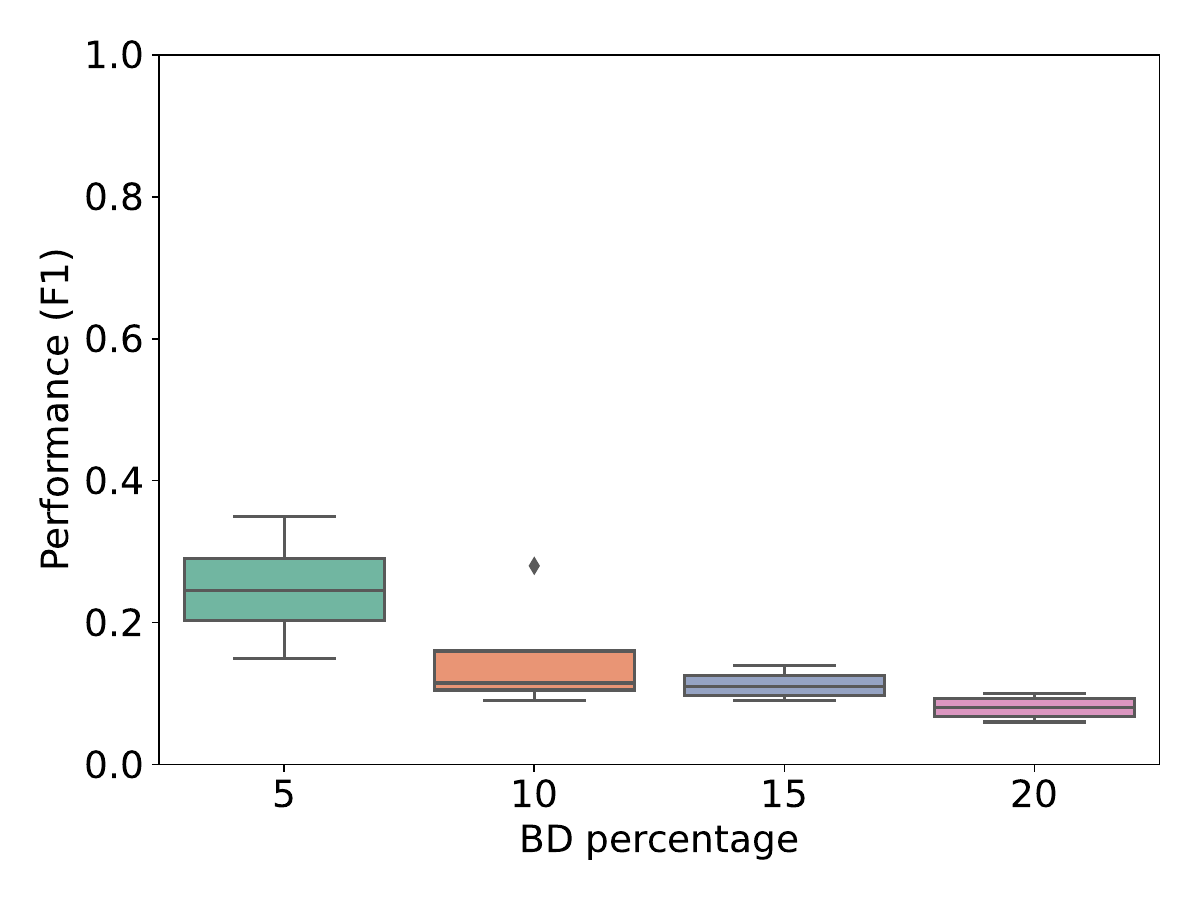}
    \caption{Backdoor performance for varying backdoor sample percentage}
    \label{fig:BD-F1}
\end{figure}

We next assess the backdoor performance by varying the number of trigger samples used in the training data. We inject a trigger only into class 1 data during training, and only into class 2 data during testing. We plot the test results for varying backdoor data used during training. We vary the bacdoor percentage of class 1 used in the training set from 5\% to 25\% and plot the classification F1-score in Figure ~\ref{fig:BD-F1}. As shown, the backdoor percentage increases, and so does misclassification. With a 20\% backdoor percentage, the classification F1 score is about 0.7, which indicates the success of the backdoor technique.

\subsection{Why Continual Learning Amplifies Backdoors in IoT}
In the case of a replay-induced reinforcement technique, replay buffers repeatedly expose poisoned samples during each update cycle, effectively re-training the backdoor over long deployment periods.
Regularization-based CL protects parameters important for past tasks, unintentionally preserving backdoor-related weights and attributing the importance parameter weight to a clean or target class.
In the case of representation lock-in, contrastive and self-supervised CL align representations over time, embedding triggers deeply into the feature space \cite{carlini2023poison,cha2021co2l}.
IoT/CPS systems rarely retain full historical data due to resource limitations, making post-hoc backdoor detection extremely difficult.

\section{Conclusion}
In this paper, we propose a backdoor training technique for an anomaly detection model in the IoT/CPS domain, using a continual learning approach.
Our experiments proves backdoor attacks represent a critical threat to continual learning in IoT systems. The same mechanisms that enable lifelong adaptation—replay, regularization, and representation reuse—also enable persistent and stealthy malicious behaviors. Addressing these challenges requires security-aware continual learning designs tailored to IoT constraints.
In the future, we plan to test the model's robustness against various defense techniques and propose an adaptive defense mechanism.

\bibliographystyle{IEEEtran}
\bibliography{bib}

\end{document}